# Comparison of Multi Criteria Decision Making Algorithms for Ranking Cloud Renderfarm Services

**J. Ruby Annette[1*], Aisha Banu[2] and P. Subash Chandran[3]**

[1]IT Department, B.S. Abdur Rahman University, Chennai - 600048, Tamil Nadu, India; Rubysubash2010@gmail.com
[2]CSE Department, B.S. Abdur Rahman University, Chennai – 600048, Tamil Nadu, India; aisha@bsauniv.ac.in
[3]Niteo Technologies, Chennai - 600017, Tamil Nadu, India; subash24@gmail.com

## Abstract

Cloud services that provide a complete environment for the animators to render their files using the resources in the cloud are called Cloud Renderfarm Services. The objective of this work is to rank and compare the performance of these services using two popular Multi Criteria Decision Making (MCDM) Algorithms namely the Analytical Hierarchical Processing (AHP) and SAW (Simple Additive Weighting) methods. The performance of three real time cloud renderfarm services are ranked and compared based on five Quality of Service (QoS) attributes that are important to these services namely the Render Node Cost, File Upload Time, Availability, Elasticity and Service Response Time. The performance of these cloud renderfarm services are ranked in four different simulations by varying the weights assigned for each QoS attribute and the ranking obtained are compared. The results show that AHP and SAW assigned similar ranks to all three cloud renderfarm services for all simulations.

## 1. Introduction

Rendering is an inevitable time consuming process in the animation industry that is usually completed using a cluster of computers in the animation studio called an Offline render farm. An Offline renderfarm is a collection of render nodes. A render node is nothing but the individual computer in the cluster in which the animation scene file that is split into individual frames is rendered in a distributed manner at the same time. In order to reduce the rendering time, new technologies like grid[1-5] and cloud computing were also explored[6-10] and found to be fruitful. Among cloud services, the PaaS type of cloud renderfarm services model is gaining popularity. The advantage of using the PaaS type of cloud renderfarm services is that the animators need no prior technical knowledge about the cloud environment and they need to pay only the render node cost for every hour of rendering. To render the files in cloud based renderfarms, the animators upload the animation files to be rendered onto the service provider server. The service Provider uses software called the Rendering Job Manager (RJM), to perform the task of a queue manager and assign virtual machines in cloud based on the scheduling policy for completing the rendering tasks. The RJM scales up or scales down the number of virtual machine instantly in order to achieve the deadline.

In this work, we compare the performance of three cloud renderfarm services using two Multi Criteria Decision Making (MCDM) Algorithms namely the Analytical Hierarchical Processing (AHP)[11,12] and SAW (Simple Additive Weighting)[13] and draw useful insights. There are many works focusing on cloud services[14-23] and the ranking of Infrastructure-as-a-Service (IaaS)

*Author for correspondence

cloud services[24]. But very few works are aboutPaaS cloud renderfarm services[25] according to our literature survey. This work bridges this gap and contributes in the following ways: a) Identifies QoS attributes specific to Cloud Renderfarm services and ranks them using AHP and SAW (Section 2). b) Compares performance of the two MCDM algorithms based on four simulation results. C) Analyses the results obtained and provide useful insights (Section 4) and finally, conclude the work with scope for further work (Section 5).

## 2. Ranking Services Using AHP and SAW

The AHP method of ranking helps the user to specify his overall goal in the form of a hierarchical diagram in order to compare the decision alternatives efficiently. The AHP method first decomposes the Renderfarm selection problem into various levels like the overall goal to be achieved, the important criteria, sub-criteria to be considered and the decision alternatives. The overall goal in this work is to rank and select the best cloud render farm service that satisfies the user Quality of Service (QoS) requirements. The five QoS attributes were selected based on Service Measurement Index (SMI metrics)suggested by Cloud Service Measurement Index Consortium for cloud services (CSMIC). The five criteria selected for ranking the cloud Render farm services are the Render node cost, File Upload Time, Availability, Elasticity and the Service Response Time (SRT).A hierarchy diagram of QoS attributes for cloud renderfarm services is designed as given Figure 1. Next a pair wise comparison and prioritization of the attributes is performed from the lower level to the top level by estimating the relative importance of attributes considered. The relative importance of an attribute can be calculated by assigning relative weights to each attribute within each level in such a way that the sum of all the relative weights of all the attributes in each level is equal to one. The next step in this process is to compute the relative ranking for all the Sub-level attributes for which a Relative Ranking Matrix of size NR x NR is formed. Where, NR is the total number of Services to be compared for ranking. Using this Relative Ranking Vector calculated from the above step, the relative ranking of all the services for one sub-level attribute called the Eigen value is estimated using the instruction given in[11] and[12]. Finally the relative ranking vector is aggregated for each sub attribute at each level and the final relative ranking vector is computed. The Final Relative Ranking vector is sorted and the sorted list of cloud renderfarm services according to their ranks is obtained. However, the five QoS attributes considered in this work do not have any sub levels.

### 2.1 SAW Method of Ranking

The Simple Additive Weighting (SAW) method also known as the WSM (Weighted Sum Model) or the weighted linear Combination method or SM (Scoring method) calculates the overall score of a cloud render farm service by calculating the weighted sum average of all the attribute values. SAW calculates an evaluation score for every alternative by multiplying the relative importance weights directly with the normalized value of the criteria for each alternative assigned by the user. The obtained product value is then summed up. The alternative service with the highest score is selected as the best cloud renderfarm service. The formula to calculate the overall score ($S_i$) for an (N) alternative service with (M) QoS attributes is given below in Equation 1.

$$S_i = \sum_{j=0}^{M} w_j r_{ij} \quad \text{(Equation 1) For } i = 1, 2, 3, \ldots N$$

Where $r_{ij}$ refers to the normalized rating, 'i' indicates the $i^{th}$ alternative and 'j' indicates the $j^{th}$ criterion. $w_j$ is the $j^{th}$ criterion weight. The formula to calculate benefit criteria value of $r_{ij}$ is given below by the equation 2.

$$r_{ij} = x_{ij} / \max_i(x_{ij}) \quad \text{(Equation 2)}$$

Similarly, the formula to calculate the worst criteria value of $r_{ij}$ is given by equation 3.

$$r_{ij} = (1/x_{ij}) / \max_i(1/x_{ij}) \quad \text{(Equation 3)}$$

Where, $x_{ij}$ represents the original value of the $j^{th}$ criterion of the $i^{th}$ alternative.

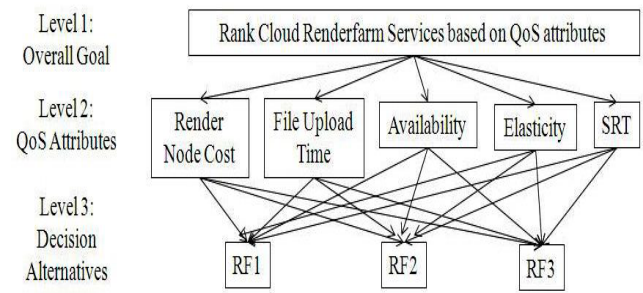

**Figure 1.** AHP Hierarchy of QoS Attributes for Cloud Renderfarm Services.

## 3. Performance Comparison of Ranking Algorithms

To compare the performances of the two algorithms four different simulations with four different criteria of importance for QoS attributes were set as given above in the Table 1. The QoS details of the five attributes selected are collected for 3 real time cloud renderfarm services. The detail regarding the service providers are not provided to avoid the hindrance that may be caused to the service providers. In the simulation 1, the two QoS attributes or criteria namely the Render Node Cost and the File Upload Time are given more preference than the other criteria as given in Table 2. In Simulation 2, the Service Response Time (SRT) attribute is given more emphasize than any other QoS attribute as given in Table 3. In Table 4, it is worthy to note that, the criteria emphasized in Simulation 1 and 3 are similar as the two criteria namely the render node cost and the File upload time are given more preference than the other criteria. In Simulation 4, the file upload time is given more importance than the other QoS attributes as given in Table 5.

## 4. Results and Discussion

The performance of three real time cloud renderfarm services are ranked and compared based on five Quality of Service (QoS) attributes that are important to these services namely the Render Node Cost, File Upload Time, Availability, Elasticity and Service Response Time. The performance of these cloud renderfarm services are ranked in four different simulations by varying the

**Table 1.** Weight Assigned to QoS Attributes for each Simulation

| Sim / QoS | Render Node Cost | FileUpload Time | Avail | Elast | SRT | CR |
|---|---|---|---|---|---|---|
| Sim1 | 0.47821 | 0.35242 | 0.04562 | 0.05432 | 0.06943 | 0.0000 |
| Sim2 | 0.24562 | 0.16293 | 0.03241 | 0.02452 | 0.53452 | 0.049 |
| Sim3 | 0.40251 | 0.30321 | 0.02254 | 0.02548 | 0.24626 | 0.049 |
| Sim4 | 0.03214 | 0.86782 | 0.01235 | 0.01253 | .07516 | 0.048 |

Abbreviations:
Sim – Simulation, Avail- Availability, Elast – Elasticity, SRT – Service Response Time,
CR- Consistency Ratio

**Table 2.** Ranking based on Simulation 1 attribute weights

|  | AHP | SAW |
|---|---|---|
| RF1 | 0.4532 Rank # 2 | 0.4752 Rank # 2 |
| RF2 | 0.8213 Rank # 1 | 0.7958 Rank # 1 |
| RF3 | 0.2132 Rank # 3 | 0.2310 Rank # 3 |

**Table 3.** Ranking based on Simulation 2 attribute weights

|  | AHP | SAW |
|---|---|---|
| RF1 | 0.8562 Rank # 1 | 0.8469 Rank # 1 |
| RF2 | 0.2354 Rank # 3 | 0.2310 Rank # 3 |
| RF3 | 0.3959 Rank # 2 | 0.4065 Rank # 2 |

**Table 4.** Ranking based on Simulation 3 attribute weights

|  | AHP | SAW |
|---|---|---|
| RF1 | 0.4852 Rank # 2 | 0.4789 Rank # 2 |
| RF2 | 0.8356 Rank # 1 | 0.8267 Rank # 1 |
| RF3 | 0.2210 Rank # 3 | 0.2198 Rank # 3 |

**Table 5.** Ranking based on Simulation 4 attribute weights

|  | AHP | SAW |
|---|---|---|
| RF1 | 0.2567 Rank # 3 | 0.2412 Rank # 3 |
| RF2 | 0.3590 Rank # 2 | 0.3502 Rank # 2 |
| RF3 | 0.7956 Rank # 1 | 0.7902 Rank # 1 |

weights assigned for each QoS attribute and the ranking obtained are compared. The AHP method based ranks for Renderfarms is given in Figure 2 and the ranking of the renderfarms using the SAW method is given in Figure 3. It can be clearly seen that, in Simulation 1 and 3, the RF2 is selected as the best alternative to achieve the criteria of low render node cost and low File upload time by both the algorithms. In simulation 2, the lowest Service Response Time criterion is best achieved by RF1 as ranked best service by both AHP and SAW methods. Since in Simulation 4, the file upload time is taken as the prime criteria the RF3 renderfarm service is selected as the best alternative with lowest file upload time by both AHP and SAW. From the above discussions, it is evident that both AHP and SAW has similar ranking for each simulation criteria discussed and the ranking values are also very close to each other.

## 5. Conclusion and Future Work

This work compared and analyzed the raking of cloud renderfarm services using two Multi Criteria Decision Making (MCDM) methods. It was observed that the ranking value differs based on the weights assigned to each QoS attribute. The best cloud renderfarm service differs based on the QoS criteria weights. However, both AHP and SAW has similar ranking for each simulation criteria discussed and the ranking values are also very close to each other. Thus SAW method is a good alternative for the AHP and could be preferred instead of AHP when no hierarchy of attributes exists as in the case discussed in this work. However, if there many hierarchy levels with sub attribute then AHP qualifies as a better method to find out the aggregated rank value. In future, the QoS attributes that are more relevant to the cloud based rendering would be identified and the ranking would be simulated with different criterion using various other MCDM algorithms and the results would be compared to identify the efficiency and significance of the MCDM methods.

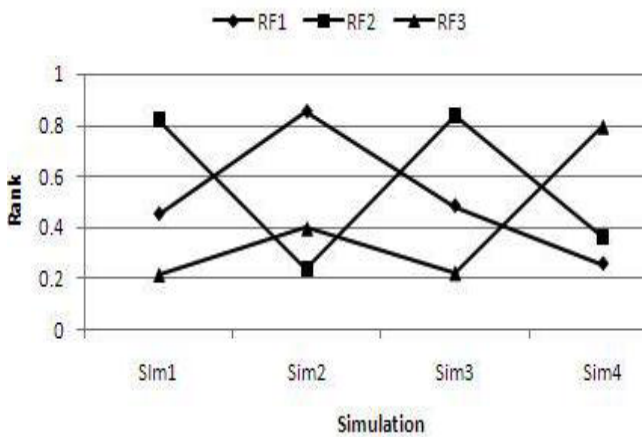

**Figure 2.** AHP Method Based Ranks for Renderfarms.

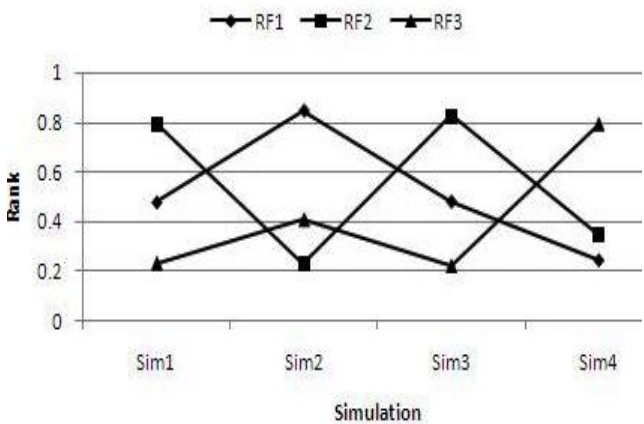

**Figure 3.** SAW Method Based Ranks for Renderfarms.

## 6. References

1. Glez-Morcillo C et al. A New Approach to Grid Computing for Distributed Rendering. 2011 International Conference on P2P, Parallel, Grid, Cloud and Internet Computing (3PGCIC), IEEE. 2011.
2. Patoli MZ et al. An open source grid based render farm for blender 3d. Power Systems Conference and Exposition, PSCE'09. IEEE/PES. IEEE. 2009.
3. Patoli Z et al. How to build an open source render farm based on desktop grid computing. Wireless Networks, Information Processing and Systems. Springer Berlin Heidelberg. 2009; 268–78.
4. Gkion M et al. Collaborative 3D digital content creation exploiting a Grid network. International Conference on Information and Communication Technologies, 2009. ICICT'09, IEEE. 2009.
5. Chong AS, Levinski K. Grid-based computer animation rendering, InGRAPHITE '06: Proceedings of the 4th International Conference on Computer Graphics and Interactive Techniques in Australasia and Southeast Asia, New York, NY, USA, ACM. 2006.
6. Baharon MR et al. Secure rendering process in cloud computing. 2013 Eleventh Annual International Conference on Privacy, Security and Trust (PST), IEEE. 2013.
7. Kennedy J, Healy P. A method of provisioning a Cloud-based renderfarm, EP2538328 A1 [Patent]

8. Cho K et al. Render Verse: Hybrid Render Farm for Cluster and Cloud Environments. 2014 7th Conference on Control and Automation (CA), IEEE. 2014.
9. Carroll MD, Hadzic I, Katsak WA. 3D Rendering in the Cloud. Bell Labs Technical Journal. 2012; 17(2):55–66.
10. Srinivasa G et al. Runtime prediction framework for CPU intensive applications, U.S. Patent No. 7,168,074. 2007.
11. Tran VX, Tsuji H, Masuda R. A new QoS ontology and its QoS-based ranking algorithm for Web services. Simulation Modelling Practice and Theory. 2009; 17(8):1378–98.
12. Saaty TL. How to make a decision: the analytic hierarchy process. European Journal of Operational Research. 1990; 48(1):9–26.
13. Stevens-Navarro E, Wong VW. Comparison between vertical handoff decision algorithms for heterogeneous wireless networks. 2006 . In Vehicular technology conference, (2006). VTC 2006-Spring. IEEE 63rd. 2006; 2:947–51.
14. Kalpana V, Meena V. Study on data storage correctness methods in mobile cloud computing. Indian Journal of Science and Technology. 2015 Mar; 8(6). Doi:10.17485/ijst/2015/v8i6/70094.
15. Durairaj M, Manimaran A. A study on security issues in cloud based e-learning. Indian Journal of Science and Technology. 2015 Apr; 8(8). Doi:10.17485/ijst/2015/v8i8/69307.
16. Shyamala K, Sunitha Rani T. An analysis on efficient resource allocation mechanisms in cloud computing. Indian Journal of Science and Technology. 2015 May; 8(9). Doi:10.17485/ijst/2015/v8i9/50180.
17. Rajasekaran A, Kumar A. User preference based environment provisioning in cloud. Indian Journal of Science and Technology. 2015 Jun; 8(11). Doi:10.17485/ijst/2015/v8i11/71781.
18. John SM, Mohamed M. Novel backfilling technique with deadlock avoidance and migration for grid workflow scheduling. Indian Journal of Science and Technology. 2015 Jun; 8(12). Doi:10.17485/ijst/2015/v8i12/60755.
19. Bagheri R, Jahanshahi M. Scheduling workflow applications on the heterogeneous cloud resources. Indian Journal of Science and Technology. 2015 Jun; 8(12). Doi:10.17485/ijst/2015/v8i12/57984.
20. Uddin M, Memon J, Alsaqour R, Shah A, Abdul Rozan MZ. Mobile agent based multi-layer security framework for cloud data centers. Indian Journal of Science and Technology. 2015 Jun; 8(12). Doi:10.17485/ijst/2015/v8i12/52923.
21. Meghana Ramya Shri J, Subramaniyaswamy V. An effective approach to rank reviews based on relevance by weighting method. Indian Journal of Science and Technology. 2015 Jun; 8(11). Doi:10.17485/ijst/2015/v8i11/61768.
22. Moayeri M, Shahvarani A, Behzadi MH, Hosseinzadeh-Lotfi F. Comparison of Fuzzy AHP and Fuzzy TOPSIS Methods for Math Teachers Selection. Indian Journal of Science and Technology. 2015 Jul; 8(13). Doi:10.17485/ijst/2015/v8i13/5410.
23. Venkatesan N, ArunmozhiArasan K, Muthukumaran S. An ID3 algorithm for performance of decision tree in predicting student's absenteeism in an academic year using categorical datasets. Indian Journal of Science and Technology. 2015 Jul; 8(14). Doi:10.17485/ijst/2015/v8i14/72730.
24. Garg SK, Versteeg S, Buyya R. SMICloud: a framework for comparing and ranking cloud services 2011. 2011 Fourth IEEE International Conference on Utility and Cloud Computing (UCC). 2011. p. 210–8.
25. Annette R, Aisha Banu W. Article: A Service Broker Model for Cloud based Render Farm Selection 2014. International Journal of Computer Applications, Published by Foundation of Computer Science, New York, USA. 2014; 96(24):11–4.